\documentclass{PoS}

\usepackage{amsmath}
\usepackage{booktabs}
\usepackage[shortcuts]{extdash}
\usepackage{wrapfig}

\title{Applicability of Quasi-Monte Carlo for lattice systems}

\ShortTitle{QMC on the lattice}

\author{\speaker{Andreas~Ammon},$^{a,b}$ Tobias~Hartung,$^{c}$ Karl~Jansen,$^b$ Hernan~Leovey,$^d$ Andreas~Griewank$^d$ and Michael~M{\"u}ller-Preussker$^a$\\
\\
$^a$Humboldt-University Berlin, Department of Physics\\Unter den Linden 6, D-10099 Berlin, Germany\\
{$^b$}NIC, DESY Zeuthen\\Platanenallee 6, D-15738 Zeuthen, Germany\\
{$^c$}King's College London, Department of Mathematics\\Strand, London WC2R 2LS, United Kingdom\\
{$^d$}Humboldt-University Berlin, Department of Mathematics\\Unter den Linden 6, D-10099 Berlin, Germany\\
E-mail: \email{Andreas.Ammon@desy.de}, \email{tobias.hartung@kcl.ac.uk}, \email{Karl.Jansen@desy.de}, \email{leovey@math.hu-berlin.de}, \email{griewank@math.hu-berlin.de}, \email{mmp@physik.hu-berlin.de}
}

\abstract{
This project investigates the applicability of quasi-Monte Carlo methods to Euclidean lattice systems in order to improve the asymptotic error scaling of observables for such theories. The error of an observable calculated by averaging over random observations generated from ordinary Monte Carlo simulations scales like $N^{-1/2}$, where $N$ is the number of observations. By means of quasi-Monte Carlo methods it is possible to improve this scaling for certain problems to $N^{-1}$, or even further if the problems are regular enough. We adapted and applied this approach  to simple systems like the quantum harmonic and anharmonic oscillator and verified an improved error scaling of all investigated observables in both cases.
}

\FullConference{31st International Symposium on Lattice Field Theory - LATTICE 2013\\
		July 29 - August 3, 2013\\
		Mainz, Germany}

\newcommand{\vx}{\mathbf{x}}
\newcommand{\vz}{\mathbf{z}}
\newcommand{\vw}{\mathbf{w}}

\renewcommand{\baselinestretch}{0.95}
\begin{document}
\section{Motivation}


The quasi-Monte Carlo (QMC) method and their randomizations (RQMC) are highly interesting for the domain of lattice field theory. It offers the possibility to improve tremendously the asymptotic error scaling of observables obtained from Monte Carlo (MC) simulations of lattice path integrals. Substantial reductions in computing time could be achieved if the QMC approach could eventually be applied to lattice-QCD (quantum chromodynamics in its lattice regularized form). A mathematical review of the QMC approach can be found in \cite{Kuo:KSS}. The major part of this contribution is based on our paper \cite{Jansen:2013jpa} (cf. also \cite{Jansen:2012xyz}). The reader interested in more details is referred to this reference at any point of the following discussion.

In order to better understand the point where the QMC approach sets in with its improvement, we want to outline the typical workflow during the treatment of a general lattice problem with conventional methods. Such a lattice system might be described by the partition function $Z = \int \mathcal{D} x  \; e^{-S[x]} $ given the action $S$. An observable $O$ could be calculated by $ \langle {O} \rangle = Z^{-1}  \int \mathcal{D} x  \; e^{-S[x]} \; O[x]$.
$\mathcal{D} x $ stands for the path integral measure of all dynamic fields relevant to the model under consideration. This could be for example the gauge field measure for lattice gauge theories or simply a particle path measure for quantum mechanical problems.
It is originated in the high dimensionality of the lattice path integral that it can naturally only be treated by means of MC methods, provided that analytic solutions are missing. Within the variety of MC approaches the Markov chain-Monte Carlo (Mc-MC) approach turns out to be the most efficient, as it allows for importance sampling.
During a Mc-MC simulation a number $N$ of field configurations $(x_i)_{i=1\ldots N}$ is generated successively, each of them based on its predecessor and distributed (after the thermalization) according to the Boltzmann weight. Then for each sample $x_i$ the observable $O$ is measured, leading to $N$ samples $O_i$. Often, these samples are distributed normal, at least to a good approximation in most cases. Then, the asymptotic error of the mean $\bar{O} = \frac{1}{N} \sum _{i=1}^N O_i$ scales like $N^{-\frac{1}{2}}$. A fixed-factor increase of the error often arises from correlations between successive observations $O_i$, which originates in the nature of Mc-MC methods.
Both features, the crude asymptotic error scaling and the possibly strong auto-correlation, lead to a necessity to generate a large amount of samples to reach a given error level. In many cases it is even impossible to reach the target accuracy due to the lack of sufficient computing resources.

The QMC approach provides the potential to circumvent the aforementioned problems, as it exhibits certain favorable properties. Most importantly, it is able to improve the error scaling to $N^{-1}$, given that certain conditions are met (see \cite{Jansen:2013jpa}).
It is further encouraging to realize that the QMC technique has already been applied successfully in other fields like financial mathematics \cite{Glasserman} for example.
But before coming to specific demonstrations of this fascinating approach, we want to take a closer look to a prominent feature of QMC samples.

\section{Quasi-Monte Carlo point sets are more uniform}
Most point sets constructed through QMC techniques fulfill a so-called low-discrepancy property (see \cite{Jansen:2013jpa}, sections 3 and 4) also referred to \emph{uniformity} or \emph{more uniform} (than conventional Monte Carlo point sets).
This property can be illustrated through a simple example in two dimensions, but whose results can be generalized to arbitrary many dimensions.

\begin{wrapfigure}{r}[0cm]{5cm}
  \includegraphics[width=5cm,page=1]{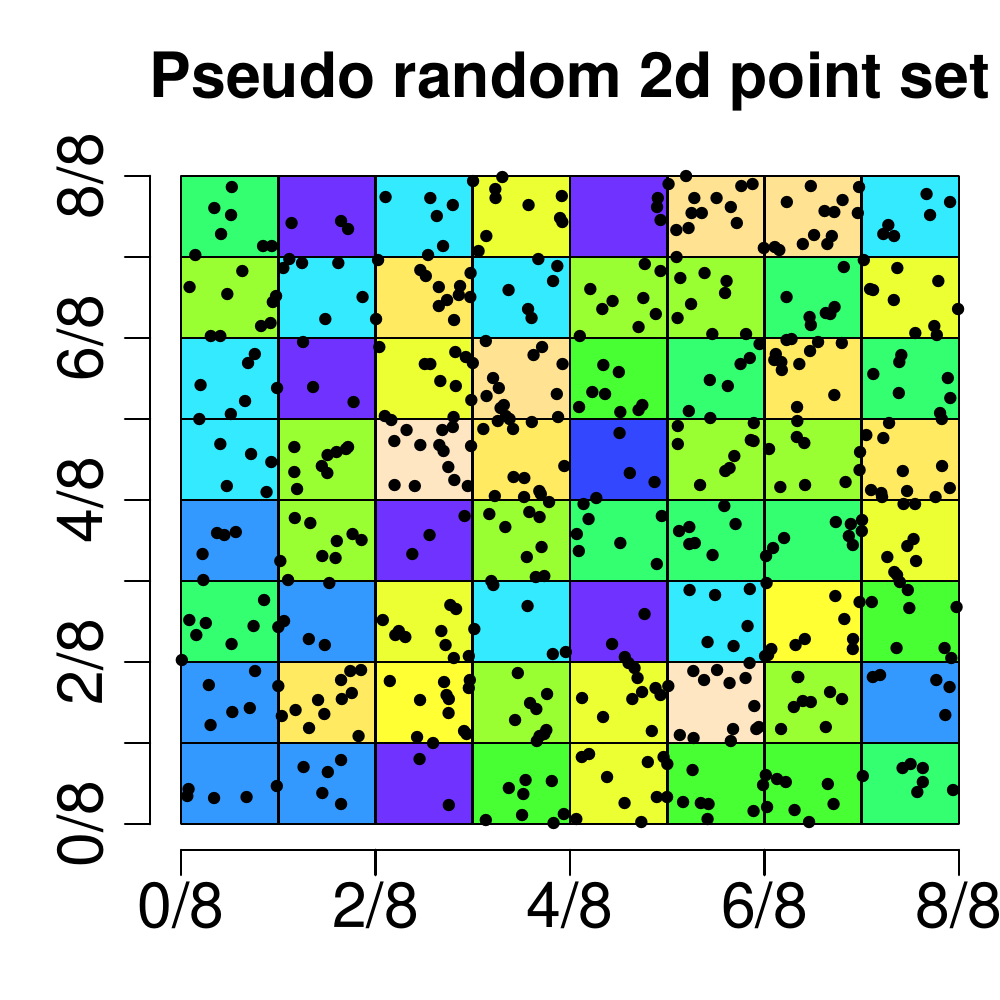}\\
  \includegraphics[width=5cm,page=2]{check_poiss_8x8_average_8}
  \caption{The pseudo-random sampling of $512$ points in a unit square (\textbf{upper plot}). For each little square the number of containing points is counted and indicated through a colour code. The meaning of each color code can be seen from the \textbf{lower diagram}, where a histogram of the counts is shown.}
  \label{fig:fig_checkerboard_mc}
\end{wrapfigure}

For this experiment the unit square $[0,1]\times[0,1]$, subdivided into $8 \times 8 $ small squares of equal size, should be considered.
Now, $512$ points are generated pseudo-randomly\footnote{We use the Mersenne Twister pseudo random number generator \cite{Matsumoto98}.} and uniformly in this unit square. Then, for each of the $8\times 8 = 64$ little squares the number of contained points are counted.
An example of the outcome of such an experiment is shown in the upper diagram of figure~\ref{fig:fig_checkerboard_mc}.

The color of each square corresponds to the number of points it contains. In the lower part of this figure we have plotted a histogram of the counts. We can clearly see that the distribution of counts is rather broad. This means in practice that many squares contain significantly more or less samples than one would expect on average, namely $8$.
If this set of points $(x_i,y_i)_{i=1\ldots 512}$ would be used for a Monte Carlo approximation of a two-dimensional integral $\int_0^1\int_0^1f(x,y)dxdy\approx\frac{1}{512}\sum_{i=1}^{512}f(x_i,y_i)$ function values in squares with very many or very few points contribute too much or too less respectively to the overall average.
This can be seen as a possibly avoidable source of extraneous fluctuations which have nothing to do with the nature of the problem, the integral of $f$.
Hence, it is highly desirable if the filling of squares with samples would happen more evenly.
We will see in the following repetition of the experiment with QMC samples, that this is exactly what can be provided by the QMC method.
We want to mention, that the distribution on the bottom of figure~\ref{fig:fig_checkerboard_mc} can be described theoretically by the Poisson distribution, in the limit of infinitely many little squares keeping the average count per little square fixed (at $8$).

\begin{wrapfigure}[24]{r}[0cm]{5cm}
  \centering
  \includegraphics[width=4.5cm,page=3]{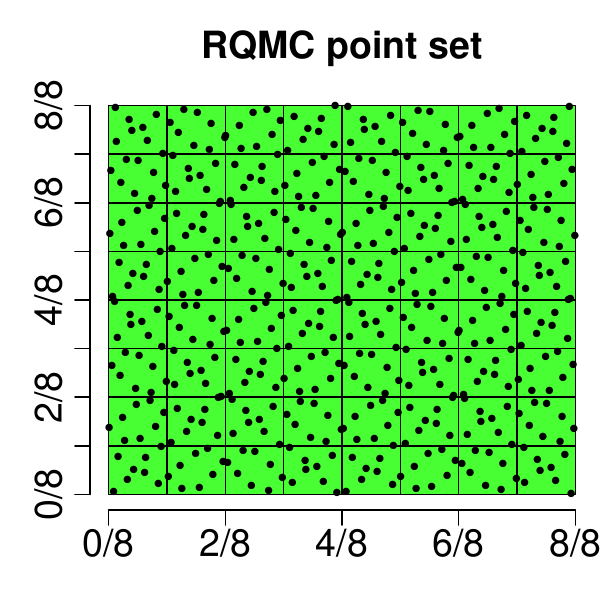}
  \includegraphics[width=4.5cm,page=4]{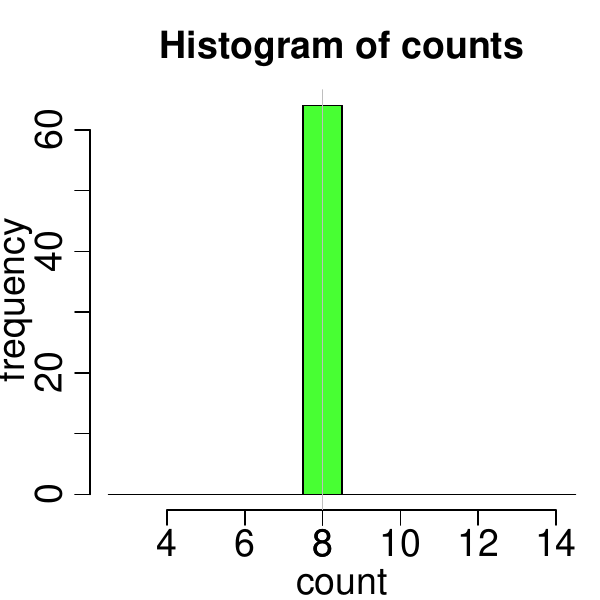}
  \caption{Distribution of $512$ Sobol' points generated uniformly in the unit square (\textbf{upper plot}) and histogram of counts (\textbf{lower plot}). See also description of figure \protect\ref{fig:fig_checkerboard_mc}.}
  \label{fig:fig_checkerboard_qmc}
\end{wrapfigure}

We repeat now the experiment with exactly the same parameters but instead of a pseudo-random number generator we employ the Sobol' approach \cite{SOB67}, a special QMC method, for the generation of points. The result is shown in figure \ref{fig:fig_checkerboard_qmc}.
As can be seen in the upper plot, the filling of squares in fact happens completely even. Each little square contains exactly $8$ points, and this leads to a delta histogram (shown on the lower part of figure \ref{fig:fig_checkerboard_qmc}).
If again the points are used in the approximation of a two-dimensional integral the function values from each square contributing to the average are represented much better with respect to the area they cover, and hence, smaller stochastic fluctuations are expected, leading very likely to smaller errors of this approximation.

Through this illustration we might get an understanding on how the more evenly distributed samples from QMC methods could help to decrease the natural statistical fluctuations of stochastic approximations.

\section{Lattice harmonic and anharmonic oscillator}
We want to briefly introduce the quantum mechanical harmonic and anharmonic oscillator quantized through the lattice path integral, which we will investigate numerically later on. These systems have been investigated in detail already in \cite{Creutz:1980gp} using the Metropolis algorithm, which is considered as a Markov chain-Monte Carlo method.

The underlying action 
\begin{equation}
  \label{eq:aho_lattice_action}
S = a \sum _ {i = 1} ^ d \left( \frac{M_0}{2} \frac{ (x_{i+1} - x_i ) ^2 }{ a^2 } + \frac{\mu^2}{2} x_i ^2 + \lambda x_i ^ 4 \right) 
\end{equation}
with the periodic boundary condition $x_{d+1} = x_1$ is obtained from the discretization of the classical mechanical action of a particle with mass $M_0$ passing along the path $x(t)$ considered in Euclidean time on an equidistant finite time lattice with lattice spacing $a$ and $d$ lattice points (extent $T=ad$). 
The time derivative $\dot{x}(t)$ is replaced by the forward finite difference $\frac{1}{a}( x_{i+1} - x_i )$. $\lambda$ controls the strength of the anharmonic term $x_i^4$. Hence, the harmonic oscillator is obtained for $\lambda = 0$ and a spring constant $\mu^2 > 0$. This condition has to be met for a convergent path integral.
The anharmonic oscillator can be simulated with $\lambda > 0$ and $\mu^2 \in \mathbb{R}$, both being finite. The case $\mu^2 < 0$, to which we restrict ourselves in the following, is particularly interesting, as the potential exhibits two minima in this case (cf. double-well potential).
The quantization is performed through the partition function $Z = \int \; e^{-S(\vx)}  \; dx_1 \ldots dx_d $. An observable $O$ of the so quantized system can be expressed as $\langle O \rangle = Z^{-1} \int \; O(\vx) \;e^{-S(\vx)}  \; dx_1 \ldots dx_d$.

The primary physical observables are
\begin{equation}
  X^2 = \frac{1}{d}\sum_{i=1}^d x_i^2,
 \; X^4 = \frac{1}{d}\sum_{i=1}^d x_i^4,
 \; \text{and} \;\;
  \label{eq:obs_correlator}
  \Gamma(\tau) = \frac{1}{d}\sum_{i=1}^d x_i x_{i+\frac{\tau}{a}} .
\end{equation}
The ground state energy $E_0 = \mu^2 X^2 + 3 \lambda X^4 + \frac{\mu^2}{16}$ and the energy gap $\Delta E = E_1 - E_0$ between the ground and first excited state can be derived from them. The latter is determined from a non-linear regression of the model $\Gamma(\tau) \sim C_0 \frac{1}{2} \left( e^{-\Delta E \tau} + e^{-\Delta E ( T - \tau ) }\right) $ to the data for the correlator $\Gamma(\tau)$, defined in \eqref{eq:obs_correlator}, in a range of times $\tau$ where the influence of higher-than-the-first excited states is negligible.

\section{Gaussian Sampling}
As the action of the harmonic oscillator is at most quadratic in the variables $x_i$, the corresponding partition function can be expressed as a multivariate Gaussian integral $Z = \int \exp\left(-\frac{1}{2} \vx ^t C^{-1} \vx\right)$, where the components of $C^{-1}$ are given by $(C^{-1})_{i,j} = \frac{2M_0}{a}\left( \left( 1+\frac{a^2 \mu^2}{2 M_0} \right) \delta_{i,j} - \frac{1}{2}\left( \delta_{i,j+1} + \delta_{i+1,j}\right) \right)$ (obtained from comparing: $\frac{1}{2} \vx ^t C^{-1} \vx = S(\vx)$). $C$ is called the covariance matrix.

For such a case, the sampling of lattice paths $\vx$ is particularly straightforward, and can be based on samples $\vz$, which are generated uniformly in the $d$-dimensional unit cube. This is particularly useful, as most RQMC methods provide samples in this form.

Hence, our algorithm aiming at the generation of properly distributed samples $\vx$ starts in the first step with
\begin{itemize}
\item[1.] the generation of a uniform sample $\vz= (z_1,\ldots,z_d)^t \in [0,1]^d$. This is either, as mentioned above, a RQMC sample stemming from a scrambled (randomized) Sobol' point set, using direction numbers from F. Kuo's page {http://web.maths.unsw.edu.au/{\textasciitilde}fkuo/sobol/index.html}, or a sample obtained from a separate uniform sampling of each dimension with a pseudo-random number generator. The latter case will be referred to as (conventional) Monte Carlo (MC) sampling in the following.
\item[2.] In the next step, univariate Gaussian samples $\vw = (w_1,\ldots,w_d)^t$ are generated by applying the inverse standard normal distribution function $\Phi^{-1}$ to the $z_i$ and multiplying the result with the square root of the eigenvalues $\lambda_i$ of $C$:
  \begin{equation}
    \label{eq:generation_wi}
    w_i = \sqrt{\lambda_i} \Phi^{-1}\left(z_{\pi^{-1}(i)}\right)\;.
  \end{equation}
The eigenvalues are given in a closed form as $\lambda_i = \left( \frac{2M_0}{a} ( u - \cos( 2\pi i /d) \right)^{-1}$. As indicated through the (inverse of the) permutation $\pi$, the order of dimensions in $\vz$ has to be modified such that the component $z_1$ comes upon the largest eigenvalue, $z_2$ comes upon the second largest eigenvalue and so on, until the last component $z_d$ meets the smallest eigenvalue. This can be achieved by determining a permutation $\pi$ which brings the eigenvalues in decreasing order ($\lambda_{\pi(1)} \leq \lambda_{\pi(2)} \leq \ldots \leq \lambda_{\pi(d)}$) and calculating $\pi^{-1}$ as the inverse of this permutation (fulfilling $\pi^{-1}(\pi(i)) = i $).
\item[3.] Finally, the multivariate Gaussian variables $x_i$ are generated from the orthonormal transformation $\vx = G \vw $, where $G = \Re(F) + \Im(F)$ is the discrete Hartley transform and $F_{k,l} = \frac{1}{\sqrt{d}}e^{2\pi ikl /d}$ the discrete Fourier transform.
\end{itemize}

\section{Results}
\begin{figure}
  \centering
  \includegraphics[width=0.495\textwidth]{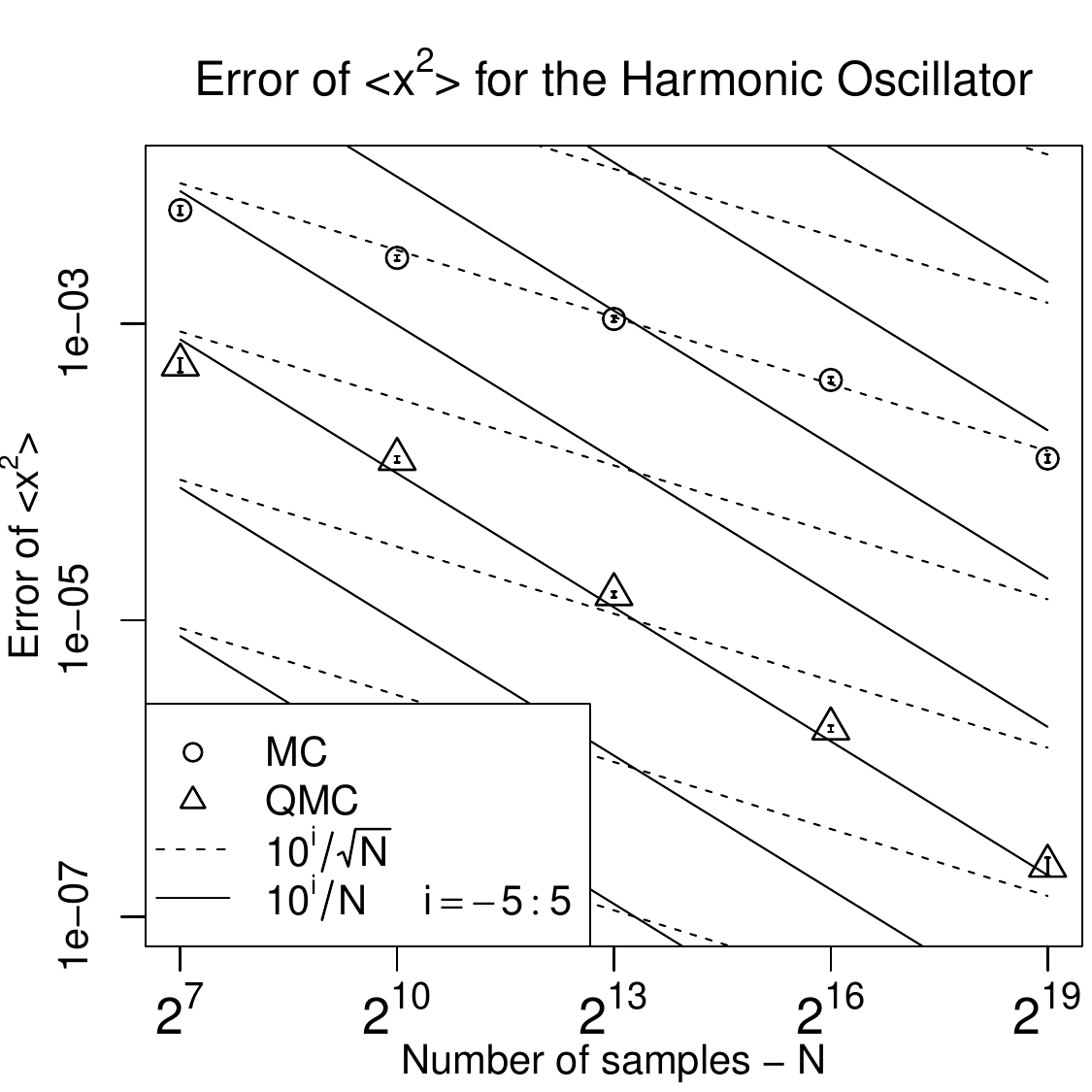}
  \includegraphics[page=2,width=0.495\textwidth]{x2_error_harm_osc_fit}
\caption{Error of $X^2$ for the RQMC and MC simulation of the harmonic oscillator (\textbf{left}). The \textbf{right} plot shows the fit of the asymptotic error scaling for the RQMC simulation.}
\label{fig:harmonic_oscillator_results}
\end{figure}

The algorithm we just discussed was used to generate lattice paths and corresponding samples of the observable $X^2$ for a harmonic oscillator with the parameters $M_0 = 0.5$, $a=0.5$, $\mu^2 = 2.0$, and $d=100$ for a fixed number of samples $N=2^{7,10,13,16,19}$. For each $N$ the experiment is repeated with $300$ scramblings (see section 5 in \cite{Jansen:2013jpa}) of the Sobol' sequence in the RQMC simulation to allow the approximation of the error and the variance of the error. For the MC simulation $300$ different seeds have been used to initialize the individual runs. A fit of the model $\Delta \langle X^2 \rangle \sim C \cdot N^\alpha$ to the determined errors of $X^2$ yields an exponent $\alpha = -1.008(15)$ for the RQMC simulation case and $\alpha = -0.49(1)$ for the MC simulation. A plot illustrating the results is shown in figure \ref{fig:harmonic_oscillator_results}. The outcome of this investigation basically proofs the full functioning of the (R)QMC method for a real, even though trivial, physical model.

\begin{figure}
  \centering
  \includegraphics[page=1,width=0.32\textwidth]{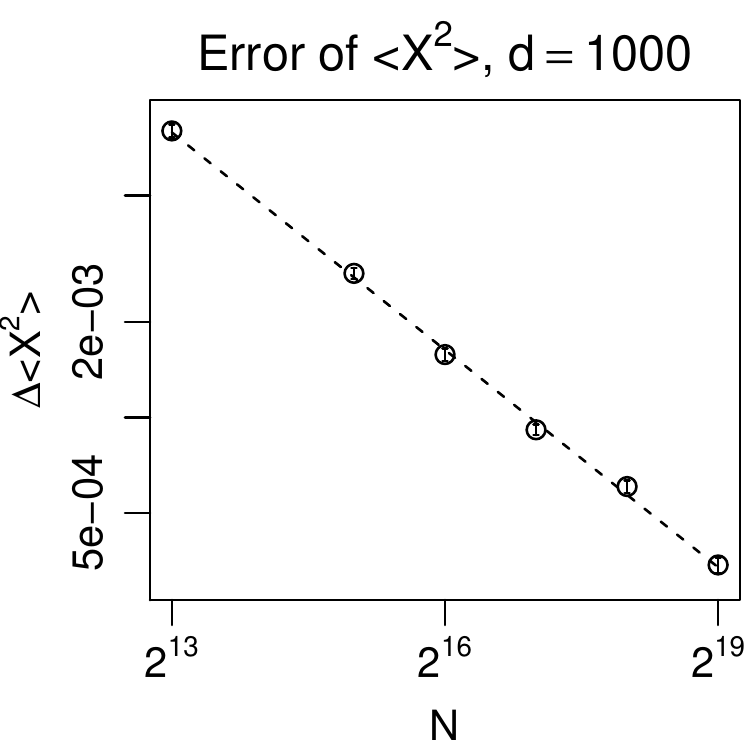}
  \includegraphics[page=2,width=0.32\textwidth]{plots_d1000_single}
  \includegraphics[page=3,width=0.32\textwidth]{plots_d1000_single}
\caption{Double-log plot of the error scaling of the observables $X^2$ (left), $X^4$ (middle) and $E_0$ (right) with the number of samples $N$ in the RQMC approach. The dashed line shows the fit of the model $\Delta O \sim C N^\alpha$ to the data. Numerical results can be seen in table \protect\ref{tab:aho_results_I}.}
\label{fig:regression_d1000}
\end{figure}

To study the scaling for a less trivial model, we passed on to the anharmonic oscillator with the parameters $\lambda = 1.0 $, $M_0=0.5$, $\mu^2 = -16$, $a=1.5 / d$. The experiment was performed for $d=100$ and $1000$ dimensions. The sampling of lattice paths is less straightforward in the present case, but can be realized on the basis of the sampling method we used before for the harmonic oscillator. This happens with the aid of the reweighting approach (cf. section 7 in \cite{Jansen:2013jpa}), which was first used in \cite{PhysRevLett_61_2635}.
Additionally to the physical action $S$, describing the anharmonic oscillator, an artificial action $S'$ is introduced, which is constructed exactly as a harmonic oscillator action with a different set of parameters $M_0'$, $a'$ and $\mu'$. Then, harmonic oscillator paths $(\vx_i)_{i=1\ldots N}$ are generated corresponding to this unphysical action $S'$. Approximations of observables $O$ of the anharmonic oscillator (described by the physical action $S$) are obtained from the weighted mean $\langle O \rangle \approx \left(\sum_{i=1}^N O(\vx_i) W(\vx_i) \right) / \left( \sum_{i=1}^N  W(\vx_i) \right)$, where the weight function $W$ is given by $W(\vx)  = \exp\left( -S(\vx) + S'(\vx) \right)$. Now, it remains to find reasonable parameters with the objective of reducing the fluctuations of the weights $W(\vx_i)$ as much as possible, leading most likely to the smallest possible error of the observables. We found that only the modification of the parameter $\mu'$ leads already to satisfying results, such that $M_0' = M_0$ and $a' = a$ could be left unchanged. A heuristic optimization approach led to a value of $\mu'=0.176$.
We adopted the procedure for the error determination as well as the regression (fits) for the exponents of the error scaling from the harmonic oscillator experiment. The results are shown in table \ref{tab:aho_results_I}.
\begin{table}[ht]
  \centering
  \begin{tabular}{ccccc}
    \toprule
          &   $O$   &  $\alpha$ & $\log C$ & $\chi^2 / \mathrm{dof} $ \\
    \midrule
            & $ X^2  $ & -0.763(8) & 2.0(1) &  7.9 / 6   \\
     $d=100$& $ X^4  $ & -0.758(8) & 4.0(1) & 13.2 / 6 \\
            & $E_0$ & -0.737(9) & 4.0(1) &  8.3 / 6 \\
    \midrule
             & $ X^2  $ & -0.758(14) & 2.0(2) & 5.0 / 4 \\
    $d=1000$ & $ X^4  $ & -0.755(14) & 4.0(2) & 5.7 / 4 \\
             & $ E_0$ & -0.737(13) & 4.0(2) & 4.0 / 4 \\
    \bottomrule    
  \end{tabular}
  \caption{Results for the error scaling of the observalbes $X^2$, $X^4$ and $E_0$ for the model of the anharmonic oscillator, simulated through reweighting. Observable errors were fitted to the model $ \Delta O \sim C N^\alpha$.}
  \label{tab:aho_results_I}
\end{table}
Plots showing the fits for $1000$ dimensions are shown in figure \ref{fig:regression_d1000}.
From table \ref{tab:aho_results_I} we can observe that the error scaling of the observables $X^2$, $X^4$ and $E_0$ is significantly improved, although less than in the harmonic oscillator case. More specifically, we can conclude that, within statistical uncertainties, the error scaling is of $O(N^{-\frac{3}{4}})$ in all considered cases.

In a further effort we investigated also the energy gap. In order to be able to measure this quantity, we increased $\mu^2$ from $-16$ to $-4$ (keeping the other parameters fixed), leading theoretically (in the limit $T\rightarrow \infty$ and $a\rightarrow 0$) to a change of $\Delta E$ from $0.0015$ to $1.576$ \cite{Blank79}. For $d=100$ dimensions and sample sizes of $N=2^{5,8,11,14}$ we obtain an exponent of $\alpha = -0.735(13)$. These results are very interesting in that the imporved error scaling seems to be rather independent of the observable under
consideration.

\section{Outlook \& conclusions}
In this contribution we showed a first successful application of one specific RQMC method to Euclidean lattice models. We verified a perfect error scaling of $O(N^{-1})$ for the harmonic oscillator and a strongly improved error scaling of $O(N^{-\frac{3}{4}})$ for the anharmonic oscillator with a double-well potential. The latter investigation also included the energy gap, which can be considered as a rather non-trivial observable, as it is obtained from  the correlator using a non-linear procedure.
A better understanding of this in-between behavior of $N^{-\frac{3}{4}}$ is planned for the future, and should lead to a better theoretical understanding on how the QMC method behaves when applied to non-trivial problems.
Further improvements in the sampling of the anharmonic oscillator are assumed when an optimally tuned, more generalized covariance matrix is used in the Gaussian sampling step.
Furthermore, other promising non-Gaussian sampling approaches, like inverse sampling \cite{Devroye:1986}, are investigated at the moment, aiming at a better description of the anharmonic action and probably involving the potential to be applicable to a much broader class of lattice problems; though, it will be interesting to see in the future how efficient these techniques are in practice.

As a next step towards lattice gauge theories we are currently considering a one-dimensional spin like model, described by the action $ S = aI \sum_{i=1}^d - \frac{1}{a^2} \cos\left(\phi_{i+1} - \phi_i \right)$, where $I$ is the moment of inertia, $a$ the lattice spacing, and $\phi_i$ are angular variables with periodic boundary conditions ($\phi_d = \phi_0$). This model exhibits topological features, visible through the non-vanishing of the winding number -- a feature not present in the previously considered oscillator models but in other lattice gauge theories like QCD.
Having managed this model with generalizable methods it could be envisaged that also generic gauge theories could be addressed in the future.

\renewcommand{\baselinestretch}{0.9}


\begin{thebibliography}{99}
\bibitem{Kuo:KSS}
  F. Kuo, C. Schwab and I. Sloan,
  \emph{ANZIAM Journal}, \textbf{53} (01) (2012).
\bibitem{Jansen:2013jpa}
  K.~Jansen, H.~Leovey, A.~Ammon, A.~Griewank, and M.~M\"uller\=/Preussker,
  \emph{Comput.Phys.Commun.} (2013), http://dx.doi.org/10.1016/j.cpc.2013.10.011, {\tt arXiv:1302.6419 [hep-lat]}.
\bibitem{Jansen:2012xyz}
  K. Jansen, H. Leovey, A. Nube, A. Griewank and M. M{\"u}ller\=/Preussker
  \emph{J. Phys.: Conf. Ser.} \textbf{454} (2013) 012043, {\tt arXiv:1211.4388 [hep-lat]}.
\bibitem{Glasserman}
  P.~Glasserman,
  Springer-Verlag, New-York 2004.
\bibitem{Matsumoto98}
  M.~Matsumoto and T.~Nishimura,
  \emph{ACM Trans. Model. Comput. Simul.,} \textbf{8} no. 1 (1998) 3--30.
\bibitem{SOB67}
  I. M. Sobol',
  \emph{U.S.S.R. Comput. Math. and Math. Phys.,} \textbf{7} no. 4 (1967) 86--112.
\bibitem{Creutz:1980gp}
  M.~Creutz and B.~Freedman,
  \emph{Annals Phys.,} \textbf{132} (1981) 427.
\bibitem{PhysRevLett_61_2635}
  A.M.~Ferrenberg and R.H.~Swendsen,
  \emph{Phys. Rev. Lett.,} \textbf{61} no. 23 (1988) 2635--2638.
\bibitem{Blank79}
  R.~Blankenbecler and T.~A.~DeGrand, Thomas A. and R.~L.~Sugar,
  \emph{Phys.Rev.,} \textbf{D21} (1980) 1055.
\bibitem{Devroye:1986}
  L.~Devroye, Springer-Verlag, New-York 1986.
\end{thebibliography}
\end{document}